\documentclass[preprint]{aastex}

\begin{document}

\title{Far-Ultraviolet Observations of RR Lyrae Stars in the Core of NGC
1851\footnote{
Based on observations made with the NASA/ESA Hubble Space Telescope,
obtained at the Space Telescope Science Institute,
which is operated by the Association of Universities for Research in
Astronomy, Inc., under NASA contract NAS 5-26555.}}

\author{Ronald A. Downes\altaffilmark{2}, 
Bruce Margon\altaffilmark{2},
Lee Homer\altaffilmark{3}, and
Scott F. Anderson\altaffilmark{3}
}

\altaffiltext{2}{Space Telescope Science Institute, 3700 San Martin Drive,
Baltimore, MD 21218; downes@stsci.edu, margon@stsci.edu}
\altaffiltext{3}{Department of Astronomy, University of Washington, Box
351580,
Seattle, WA 98195-1580; anderson@astro.washington.edu, homer@astro.washington.edu}

\begin{abstract}

There are extraordinarily few far ultraviolet observations of RR~Lyrae
stars in the literature. We present {\it Hubble Space Telescope}~FUV
($\sim1600$\AA) imaging of the core of the globular cluster NGC
1851. Eleven new variables whose light curves are consistent with
those of RR~Lyr stars are discovered, increasing the total number of
RR~Lyr known in this cluster by a substantial amount. In agreement
with basic physical theory, the observed amplitude of the variables in
the FUV is enormous compared with the century of past optical
observations, ranging up to 4~mag. {\it HST} STIS FUV observations of
cluster cores may prove an especially effective means of obtaining a
near-complete census of RR~Lyr stars, combining high angular
resolution, suppression of luminous red stars, and data where the
stellar pulsation amplitudes are greatly enhanced. Attention is also
drawn to a peculiar blue object in the cluster that is most probably a
low mass x-ray binary system in quiescence.

\end{abstract}


\section{Introduction}

A complete census of RR~Lyrae stars in a globular cluster is
important, as it is the basis for determining the Oosterhoff type of
the cluster, which in turn is used to sort various cluster properties.
\citet{hog73} first identified 10 variables in NGC~1851, while
\citet{lil75} classified and obtained periods for those objects, plus
an additional four variables.  Eleven of the 14 variables were
identified as RR~Lyr types, and based on the mean period of the
RR{\it ab} ($<0.6$~days), \citet{lil75} classified the cluster as
Oosterhoff Type~I.  \citet{wldc78} further refined the periods for the
twelve previously known objects (one of the variables for which
\citet{lil75} failed to detect any variations is in fact an RR~Lyr
star), as well as determined the periods of an additional seven
objects.  Those authors found the mean period of the RR{\it ab} was
0.573~days, consistent with \citet{lil75}, and the ratio of RR{\it
c}:RR{\it ab}~=~0.36.  Thus, NGC~1851 was at the extreme end of the
Oosterhoff Type~I clusters listed in \citet{wd77}.

\citet{wlcw82} discovered three more RR~Lyr variables, raising the
total to 22 objects, and found that the mean period of the RR{\it ab}
was now 0.572~days.  However, the ratio of RR{\it c}:RR{\it ab} jumped
dramatically from 0.36 to 0.47, again putting NGC~1851 at the extreme
end of Oosterhoff Type~I clusters.  More recently, \citet{wal98}
performed a thorough search of NGC~1851 for RR~Lyr stars in the
optical, and found seven more variables for a total of 29.  The mean
period of the RR{\it ab} increased slightly to 0.586~days, but the
ratio of RR{\it c}:RR{\it ab} was reduced back to 0.38.  \citet{ccc03}
list basic data on globular clusters with at least 10 RR{\it ab} and
5 RR{\it c} variables, and NGC~1851 has the most extreme mean period
of the RR{\it ab}, while the ratio RR{\it c}:RR{\it ab} is high but not
extreme.  However, if the census of RR~Lyr stars is incomplete, the
true standing of the cluster with respect to other Oosterhoff Type~I
clusters (which may also be incomplete) is uncertain.  

As noted by \citet{wal98}, RR~Lyr stars in the central region of the
cluster, which is too crowded for ground-based observations to
resolve, are certainly expected. Recently \citet{sum04} have reported
19 further variables in the cluster, some quite close to the core,
using ground-based data, and they conclude that most are RR~Lyr
stars. Also, \citet{sav98} used pre-COSTAR {\it Hubble Space Telescope
(HST)} WF/PC data to identify 7 new RR~Lyr candidates.  It is clear
that {\it HST} observations of the cluster (particularly with
instruments that correct for spherical abberation) are capable of
adding important new data on variables in the cluster center.

We obtained far-ultraviolet imaging observations of NGC~1851 with the
{\it HST}~ STIS to search for variability of the x-ray burster located
in the cluster core.  While that candidate object \citep{damd96} was
not found to be variable \citep{hamdm01}, several other objects are
easily seen (by visual inspection alone) to be highly variable, with
amplitudes of 2-4 magnitudes!  We therefore measured the light
curves of all objects in the cluster to identify all large-amplitude
variables, and examination of these light curves reveal that the
large-amplitude variables are very likely RR~Lyr stars. Thus, we may
now have a near-complete census of the NGC~1851 RR~Lyr population,
which will allow us to determine if it is a typical or atypical
Oosterhoff Type~I cluster.

\section{Observations and Analysis}

\begin{figure}
\epsscale{0.6}
\includegraphics[scale=0.4]{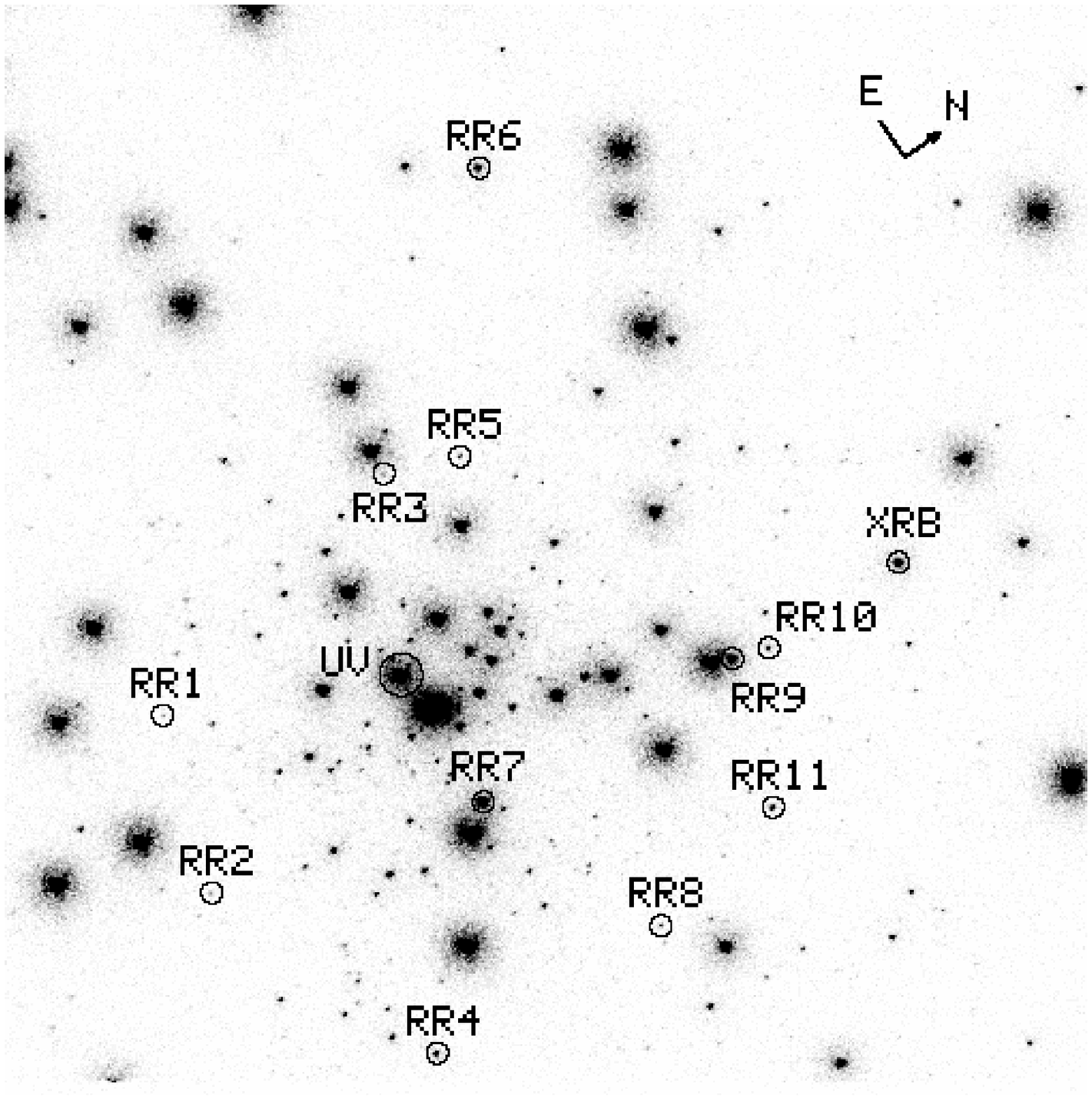}
\includegraphics[scale=0.4]{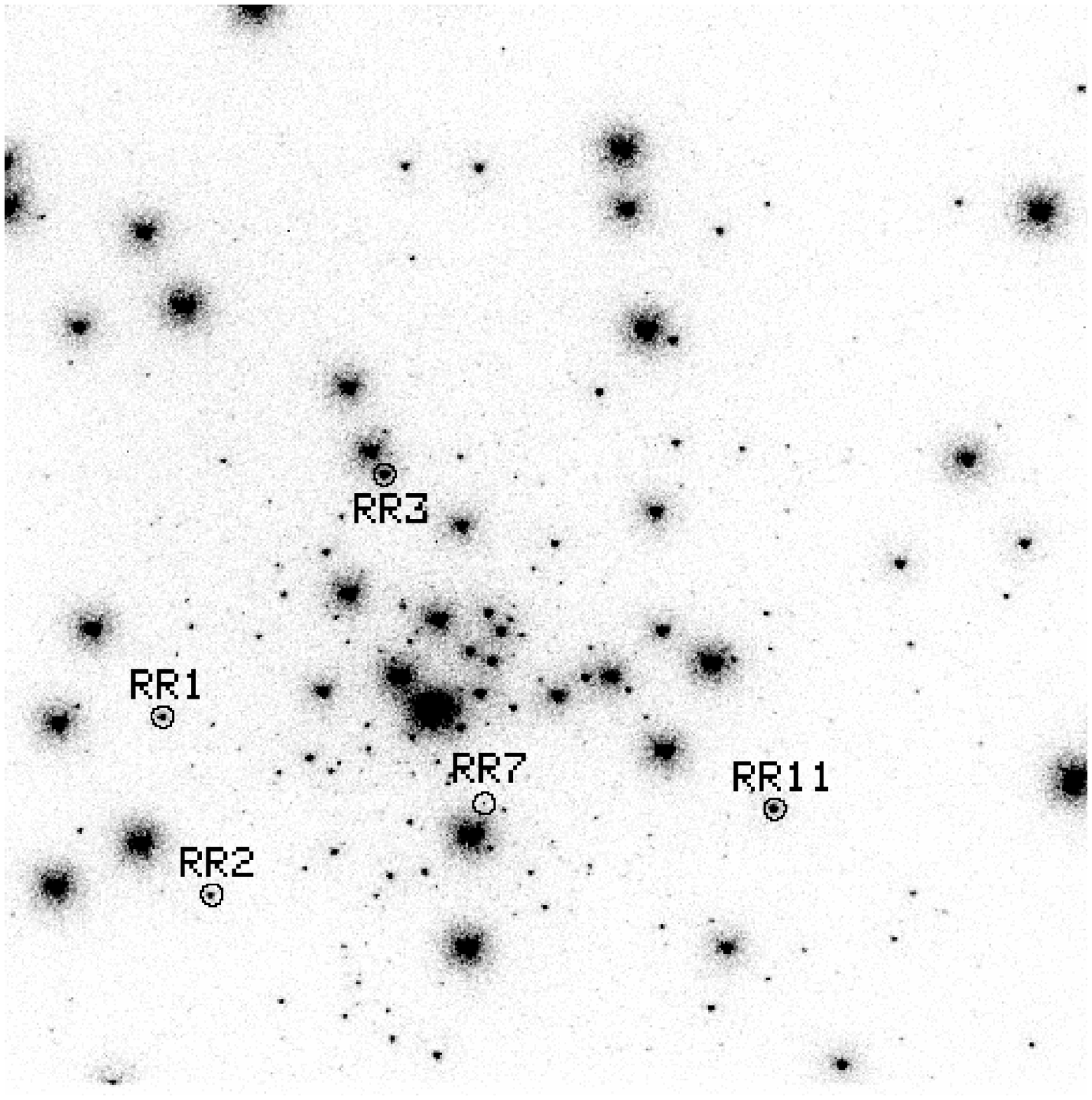}
\includegraphics[scale=0.4]{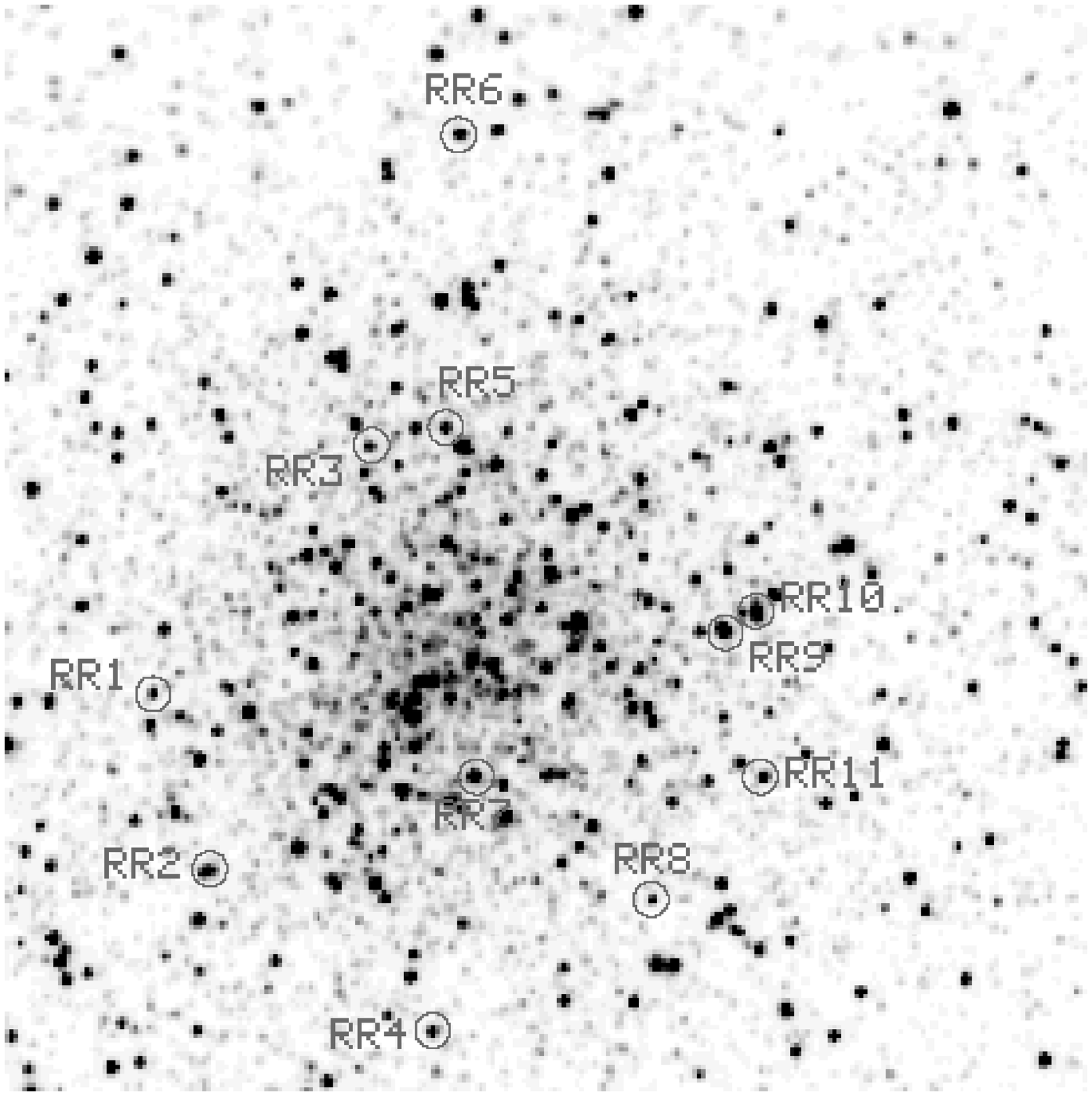}
\caption{{\it Top panels}: $25'' \times 25''$ {\it HST} STIS FUV-MAMA
images of the core of NGC 1851 with the F25QTZ filter.  The candidate
RR Lyrae stars are marked, as are two other interesting objects (see text
for details).  Objects RR1, RR2, RR3, and RR11 are at maximum on the
right image, while RR7 is at maximum on the left image.  {\it Bottom
panel}: the same field with the {\it HST} WFPC2 PC and the F439W
filter.
\label{fig1}}

\end{figure}

\begin{figure}
\epsscale{0.6}
\plotone{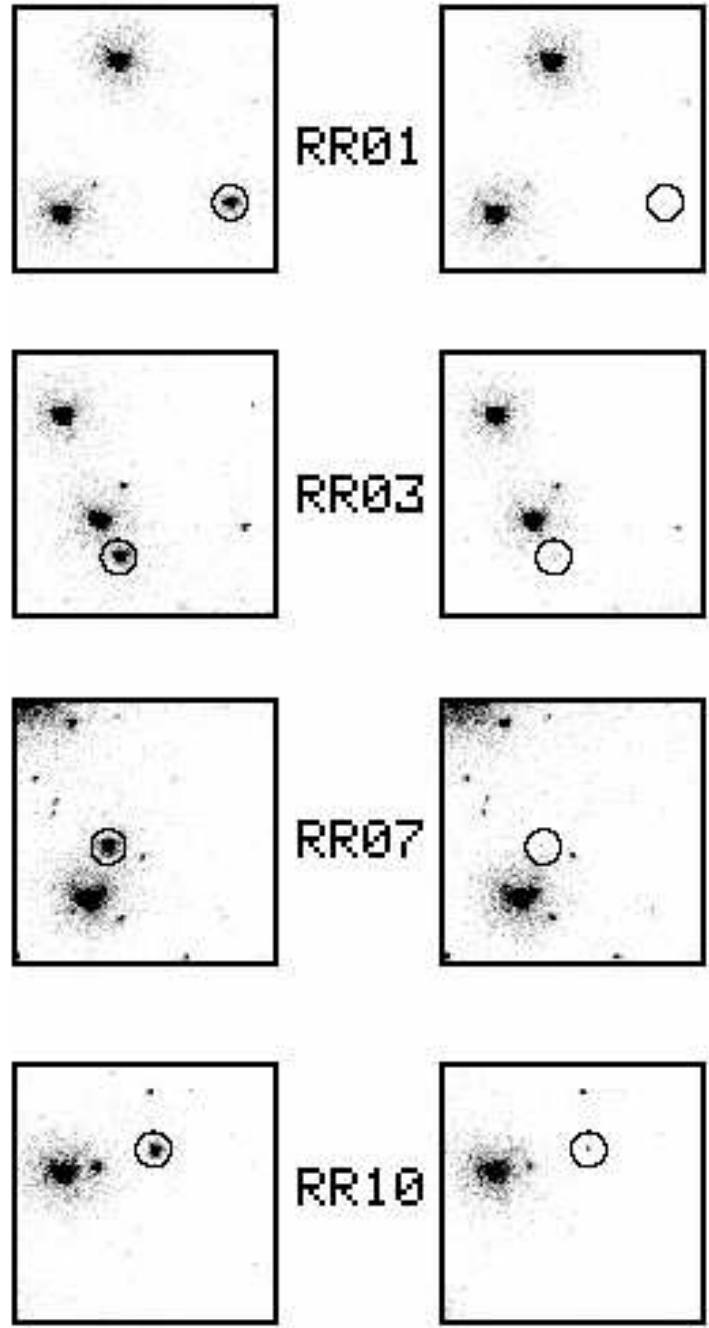}
\caption{Observations at maximum (left) and minimum (right) for a sample
of the RR Lyr stars. These images demonstrate the how straightforward
it is to detect these variables in the far-ultraviolet.
\label{fig2}}

\end{figure}

On 1999 March 24, we obtained 4 orbits of {\it HST} STIS FUV-MAMA
imaging data with the long-pass quartz (F25QTZ) filter.  This filter
has a central wavelength of 1595\AA, and a FWHM of 230\AA.  The data
were obtained in time-tag mode to support the temporal analysis
desired for the main goal of the program - a study of the x-ray
burster in the core of the cluster \citep{hamdm01}.  A thorough
analysis of the data reveals 11 objects that were significantly
variable, and basic information of these stars is given in
Table~\ref{tbl-1}. Figure~\ref{fig1} shows two of the STIS images,
with the variables identified.  To better illustrate the dramatic
nature of the variations, we show in Figure~\ref{fig2} the full range
for the four largest amplitude variables.

On 1996 April 10, we obtained {\it HST} WFPC2 images of NGC~1851
through the F439W and F336W filters; these filters are very similar to
the Johnson B and U, respectively.  To aid in identifying the objects
for future studies, we include in Figure~\ref{fig1} the F439W image.
Our prior photometric analysis \citep{damd96} determined the WFPC2
magnitudes for 7 of the variables, and these data are included in
Table~\ref{tbl-1}; the four missing objects include one not on the PC
image, and three that are in regions too crowded to obtain
accurate measurements.

Prediscovery photometry of these variables from {\it HST} is also
available in the F439W and F555W filters, approximating $B$ and $V$,
from the work of \citet{pio02}.  We also include these colors in
Table~1. It is immediately apparent that all of these stars lie in the
region of the cluster color-magnitude diagram occupied by RR~Lyr
stars. Although we have insufficient data to derive mean magnitudes,
all of the candidate objects have 0 $< (B-V) < 0.5$, in common with
the well-studied RR~Lyr's in NGC1851 discussed by \citet{wal98}. All
of our candidates also lie within 0.8 mag of the mean $V$ of the cluster
RR~Lyr's observed by \citet{wal98}, suggesting that the luminosity of
the HST variables is also consistent with this classification,
although better observations are surely desirable.

\begin{figure}

\epsscale{0.6}
\plotone{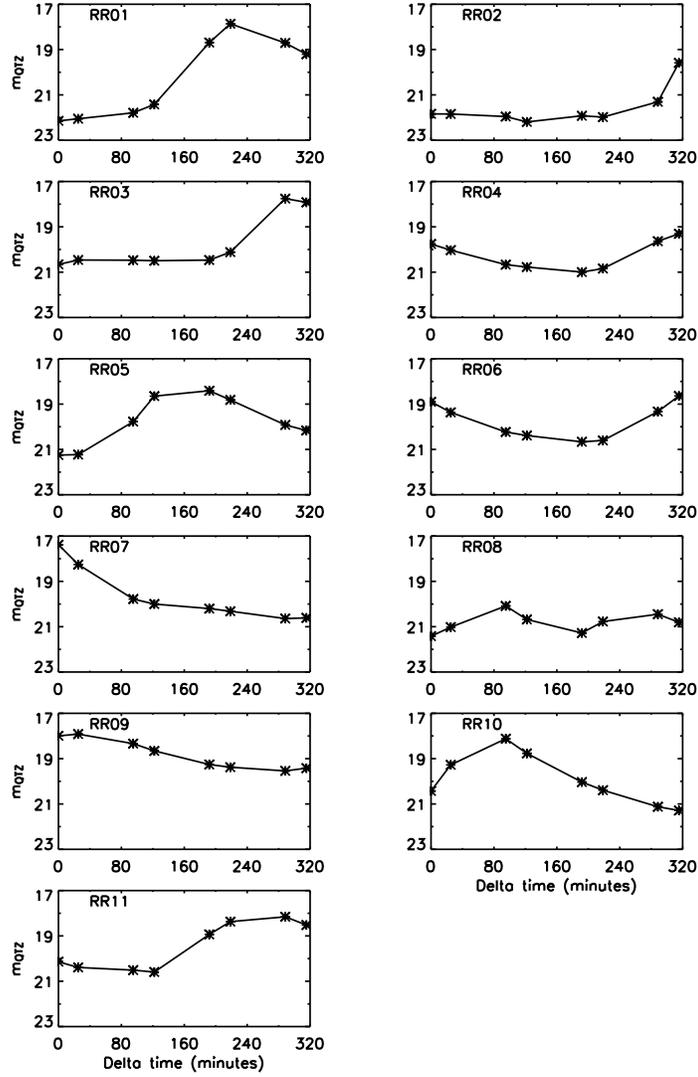}
\caption{Light curves for the 11 RR Lyrae candidates.
\label{fig3}}

\end{figure}

The light curves for the variables are shown in Figure~\ref{fig3},
which reveals that the objects are indeed most likely RR~Lyr
variables. Even though we do not cover a full period for any of the
objects, the distinctive shape of the light curves allows us to
tentatively classify most of the objects.  Note that {\it IUE}
observations of the RRab stars RR~Lyr and X~Ari \citep{bb85}, and the
RRc star DH~Peg \citep{fsjl90}, show that the light curve shapes seen
in the visible are reproduced in the ultraviolet, so FUV light curve
shape can still be used to classify the stars.  One exception may be
RR08, for which a period of $\sim160$ minutes may be appropriate, thus
ruling out an RR~Lyr classification.  It is interesting to note that
of the $\sim160$ objects detected in the STIS far-ultraviolet (FUV)
image, $7\%$ are RR~Lyr stars.  Given this high fraction, and the
extremely large amplitude, it appears that the FUV may be the best
place to search for RR~Lyr variables.

We have unsuccessfully attempted to correlate the objects discussed by
\citet{sum04} with the variables observed by {\it HST}. Nine of their
objects fall in the region covered by our data, but only six of these
have plausible counterparts in WFPC2 data (even allowing for reference
frame offsets).  Further, of those six, three are far too faint to
correspond to RR~Lyr stars in the cluster, and none appear in our FUV
data, quite contrary to our expectations.  While the correct
correlation of the \citet{sum04} objects with our {\it HST} objects
may require further observations, it is clear from the excellent
quality of the \citet{sum04} light curves that their objects are in
fact genuine RR~Lyrae stars.  Also note that none of the WF/PC
candidates of \citet{sav98} fall in our STIS field, and that Sumeral
object NV17 is coincident with Saviane object 1 (although the
classifications differ).

A search of the {\it HST} archive reveals that there is only one other
long duration, FUV imaging observation of a globular cluster that
could be used to search for RR~Lyr stars.  Analysis of this dataset on
NGC~6752 (obtained by M. Shara on 2001 March 30) revealed no RR~Lyr
stars, despite the fact that the objects should be brighter than those
in NGC~1851, as the former cluster is $\sim2$ mag closer.  A search of the
literature shows that no RR~Lyr stars have been detected in the
optical for this cluster \citep{cmd01}, so it is perhaps not
surprising that none were detected in the FUV.  

\section{Discussion}
\subsection{RR Lyrae Stars}

In the optical, RR~Lyr stars have a typical amplitude of
0.5-1.0~mag, whereas our ultraviolet data shows a larger 2-4~mag
variation.  Ultraviolet observations from the {\it ANS} satellite
of RR~Lyr itself \citep{bwbh82}
show that the amplitude of the pulsations steadily increases toward
the far-ultraviolet, going from 0.8 mag at 3300\AA\ to 3.1 mag at
$1800$\AA.  As a first approximation, the amplitude of the pulsation
should vary as (dB/dT)/B, where B is the Planck function and T is the
temperature.  Of course, this model is accurate only if the pulsation
is non-radial, which is not the case for RR~Lyr stars.  However, if
the radial variations are not a function of wavelength, then this
simplified model will be a reasonable approximation for the wavelength
dependence of the amplitude.

For RR~Lyr itself (T $\sim7000$K), we find the ratio of the amplitudes at
1800\AA~and 5500\AA~to be 3.4, so for a $\sim1$~mag amplitude at 5500\AA, this
simple model would predict an 1800\AA~amplitude of 3.4~mag, very close
to what is observed.  For our more ultraviolet wavelength (1400\AA), we
derive an amplitude of 4.6~mag.  We show, in Table~\ref{tbl-2}, the
FUV amplitudes for several temperatures for two different V amplitudes.
The ultraviolet amplitudes of the NGC~1851 RR~Lyr stars fit nicely in
these ranges.

If we consider only the well-studied objects outside the cluster core,
there were 29 previously known RR~Lyr variables, so
with the 11 new candidates presented here, the total count is now 40
variables.  While we do not have periods for the new objects, we have
determined their variability classes.  As previously noted,
\citet{wldc78} found the ratio RR{\it c}:RR{\it ab} to be 0.36,
\citet{wlcw82} increased the value to 0.47, and \citet{wal98} reduced
the value back to 0.38.  Our updated value for the ratio is 0.54,
which is the second highest of all Oosterhoff Type~I clusters
\citep{ccc03}.  If the \citet{sav98} candidates are included, the
ratio drops back to 0.47.  We do not include the candidates of
\citet{sum04} because of the concerns previously noted.  When coupled
with the fact that NGC~1851 has the most extreme mean period of the
RR{\it ab}, this cluster is truly an unusual Oosterhoff Type~I object.

\subsection{Other Interesting Objects}

\begin{figure}
\epsscale{0.6}
\plotone{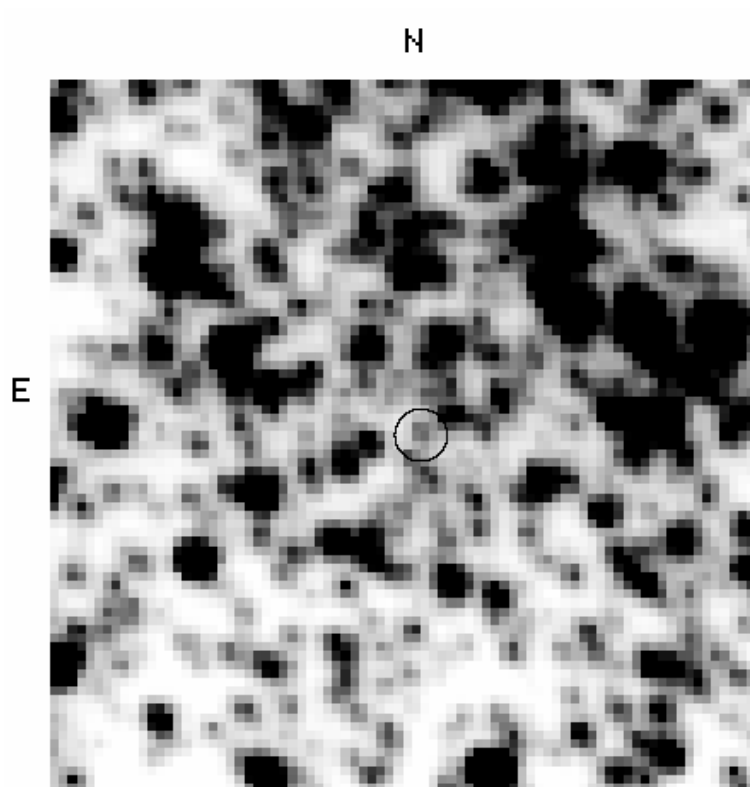}
\caption{$2'' \times 2''$ {\it HST} WFPC2 image of the UV-excess object with
the F439W filter.
\label{fig4}}

\end{figure}

In addition to looking for variable objects in the data, we combined
the STIS and WFPC2 data to search for UV-bright objects.  As noted by
\citet{damd96}, the WFPC2 data revealed four UV-bright objects, two of
which are also in the STIS image; both objects are marked in
Figure~\ref{fig1} on the STIS images.  A WFPC2 chart for the x-ray
burster can be found in \citet{damd96}, while Figure~\ref{fig4} shows
a finding chart for the other UV-bright object.  Data for these
objects --- the x-ray burster (XRB) and the other UV-bright star (UV)
--- are given in Table~\ref{tbl-3}.  Neither object is variable, while
the UV-bright star is significantly bluer in the far ultraviolet
than the x-ray burster.

Reanalysis of the {\it Chandra} HRC-S zeroth order image of NGC~1851
\citep{hamdm01} finds a possible faint X-ray source positionally
coincident with the UV-bright object.  At L$_{x} \sim10^{33}$ ergs
s$^{-1}$, this would suggest a quiescent low-mass x-ray binary
(qLMXB).  However, its optical/UV brightness (M$_{V} \sim3$) implies a
much lower L$_{x}$/L$_{opt}$ ratio than the optically identified (in
quiescence) qLMXB X5 in 47~Tucanae (L$_{x}=9.2\times10^{32}$ ergs s$^{-1}$
and M$_{V}$=8.2; \citet{ehg02}) Of course, since the {\it
Chandra} and {\it HST} observations were taken nine months apart, it
is possible that the object was in an active state during the {\it
HST} observations and a quiescent state during the {\it Chandra}
observations, which would skew the L$_{x}$/L$_{opt}$ ratio.

\section{Summary}

We have presented one of the few FUV observations of a globular cluster core, and find that RR~Lyr stars are surprising numerous and prominent in NGC~1851.
FUV observations of cluster cores may prove an especially effective means of obtaining a near-complete census of RR~Lyr stars, combining high angular resolution, suppression of luminous red stars, and data where the stellar pulsation amplitudes are greatly enhanced.

\acknowledgments

We thank E. Deutsch for providing WFPC2 photometry, and Mike Corwin
for providing coordinate information necessary to correlate the
\citet{sum04} data with the WFPC2 data.  Alister Walker provided
helpful comments on an early draft of this paper, and Howard Bond
provided software to analyze the wavelength dependence of the
pulsation amplitude.






\clearpage

\clearpage

\begin{deluxetable}{rccccccc}
\tablenum{1} \footnotesize
\tablecaption{RR Lyrae Candidates in NGC 1851\label{tbl-1}}
\tablewidth{0pt}
\tablehead{
\colhead{Name} & \colhead{R.A. (2000)\tablenotemark{a}} & 
\colhead{Decl. (2000)\tablenotemark{a}} & 
\colhead{{\it m$_{555}$}\tablenotemark{b}} & 
\colhead{{\it m$_{439}$-m$_{555}$}\tablenotemark{c}} & 
\colhead{{\it m$_{336}$-m$_{439}$}\tablenotemark{d}} & 
\colhead{Range ({\it m$_{qtz}$})\tablenotemark{e}}& \colhead{Type} 
}
\startdata

RR1  & 05 14 07.057 & -40 02 54.99 & 16.09 & 0.38 & 0.78 & 4.3 & RR{\it ab} \\
\smallskip
RR2  & 05 14 06.709 & -40 02 56.45 & 16.40 & 0.42 &      & 2.6 & RR{\it ab} \\
\smallskip
RR3  & 05 14 07.205 & -40 02 47.64 & 16.41 & 0.44 & 0.91 & 2.9 & RR{\it ab} \\
\smallskip
RR4  & 05 14 06.185 & -40 02 54.24 & 16.23 & 0.31 & 1.13 & 1.7 & RR{\it c} \\
\smallskip
RR5  & 05 14 07.147 & -40 02 45.94 & 16.29 & 0.21 & 0.89 & 2.8 & RR{\it c} \\
\smallskip
RR6  & 05 14 07.604 & -40 02 41.79 &       &      &      & 2.0 & RR{\it c} \\
\smallskip
RR7  & 05 14 06.548 & -40 02 50.05 & 15.19 & 0.06 & 0.35 & 3.3 & RR{\it c} \\
\smallskip
RR8  & 05 14 06.140 & -40 02 48.28 & 16.81 & 0.34 & 1.10 & 1.3 & RR{\it c}:: \\
\smallskip
RR9  & 05 14 06.500 & -40 02 43.43 & 16.19 & 0.22 &      & 1.6 & RR{\it c}: \\
\smallskip
RR10 & 05 14 06.476 & -40 02 42.59 & 15.42 & 0.13 &      & 3.2 & RR{\it ab} \\
\smallskip
RR11 & 05 14 06.208 & -40 02 44.61 & 16.09 & 0.17 & 0.89 & 2.4 & RR{\it ab} \\

\enddata

\tablenotetext{a}{~~Units of right ascension are hours, minutes, and seconds, 
and units of declination are degrees, arcminutes, and arcseconds. Coordinates
measured on STIS FUV-MAMA image 04wm02euq.}
\tablenotetext{b}{~~WFPC2 F555W filter}
\tablenotetext{c}{~~WFPC2 F439W and F555W filters.  Note that all 
exposures were obtained within a 15 minute interval, so variability of the 
objects should not significantly distort the color.}
\tablenotetext{d}{~~WFPC2 F336W and F439W filters.  Note that the
F336W and F439W exposures were obtained 80 minutes apart, so variability 
of the objects could distort the color. Also, this color was obtained at
a different time from the other WFPC2 values in this table.}
\tablenotetext{e}{~~STIS F25QTZ filter}

\end{deluxetable}

\clearpage

\begin{deluxetable}{rccc}
\tablenum{2}
\footnotesize
\tablecaption{Predicted Ultraviolet (1500\AA)~Amplitudes for RR Lyrae Stars\label{tbl-2}}
\tablewidth{0pt}
\tablehead{
\colhead{Temperature} & 
\colhead{UV Amplitude} & 
\colhead{UV Amplitude}  \\
\colhead{(K)} & 
\colhead{(V amp. = 0.5)} &
\colhead{(V amp. = 1.0)} 
}

\startdata

7000   & 2.3 & 4.6 \\
8000   & 1.9 & 3.8 \\
9000   & 1.7 & 3.3 \\
10000  & 1.4 & 2.8 \\

\enddata

\end{deluxetable}

\clearpage

\begin{deluxetable}{rcccccc}
\tablenum{3}
\footnotesize
\tablecaption{UV-bright Objects in NGC 1851\label{tbl-3}}
\tablewidth{0pt}
\tablehead{
\colhead{Name} & \colhead{R.A. (2000)\tablenotemark{a}} & 
\colhead{Decl. (2000)\tablenotemark{a}} & 
\colhead{{\it m$_{439}$}\tablenotemark{b}} & 
\colhead{{\it m$_{336}$-m$_{439}$}\tablenotemark{b,c}} & 
\colhead{{\it m$_{qtz}$}\tablenotemark{d}} &
\colhead{{\it m$_{qtz}$-m$_{336}$}}
}
\startdata

XRB\tablenotemark{e}  & 05 14 06.469 & -40 02 39.01 & 20.46 & -0.74 & 18.2 & -1.52 \\
\smallskip
UV\tablenotemark{f}  & 05 14 06.853 & -40 02 49.96 & 18.96 & -0.78 & 16.1 & -2.08 \\

\enddata

\tablenotetext{a}{~~Units of right ascension are hours, minutes, and seconds, 
and units of declination are degrees, arcminutes, and arcseconds. Coordinates
measured on STIS FUV-MAMA image 04wm02euq.}
\tablenotetext{b}{~~WFPC2 F439W filter}
\tablenotetext{c}{~~WFPC2 F336W filter}
\tablenotetext{d}{~~STIS F25QTZ filter}
\tablenotetext{e}{~~X-ray burster, Deutsch et al (1996)}
\tablenotetext{f}{~~UV-excess object, this work; corresponds to a faint x-ray source (see text)}
\end{deluxetable}

\clearpage

\end{document}